\documentclass[english,aps,twocolumn,prb]{revtex4}
\usepackage{textcomp}
\usepackage{bbding}
\usepackage{amsmath}
\usepackage{amssymb}
\usepackage{graphicx}
\usepackage{esint}

\makeatletter

\DeclareRobustCommand{\greektext}{%
  \fontencoding{LGR}\selectfont\def\encodingdefault{LGR}}
\DeclareRobustCommand{\textgreek}[1]{\leavevmode{\greektext #1}}
\DeclareFontEncoding{LGR}{}{}
\DeclareTextSymbol{\~}{LGR}{126}

\AtBeginDocument{\DeclareFontEncoding{T2A}{}{}}

\@ifundefined{textcolor}{}
{%
 \definecolor{BLACK}{gray}{0}
 \definecolor{WHITE}{gray}{1}
 \definecolor{RED}{rgb}{1,0,0}
 \definecolor{GREEN}{rgb}{0,1,0}
 \definecolor{BLUE}{rgb}{0,0,1}
 \definecolor{CYAN}{cmyk}{1,0,0,0}
 \definecolor{MAGENTA}{cmyk}{0,1,0,0}
 \definecolor{YELLOW}{cmyk}{0,0,1,0}
}

\makeatother

\usepackage{babel}
\begin{document}

\title{Two-dimensional Fermionic Hong-Ou-Mandel Interference with
Weyl Fermions}

\author{M. A. Khan$^{1,2,3}$, Michael N. Leuenberger$^{1,2}$}

\affiliation{$^{1}$NanoScience Technology Center, University of Central Florida,
Orlando, Florida 32826, USA}

\affiliation{$^{2}$Department of Physics, University of Central Florida, Orlando,
Florida 32816, USA.}

\affiliation{$^{3}$Federal Urdu University of Arts, Science and Technology, Islamabad,
Pakistan.}
\begin{abstract}
We propose a two-dimensional Hong-Ou-Mandel (HOM) type interference experiment
for Weyl fermions in graphene and 3D topological
insulators. Since Weyl fermions exhibit linear dispersion, similar
to photons in vacuum, they can be used to obtain the HOM
interference intensity pattern as a function of the delay time between
two Weyl fermions. We show that while the Coulomb interaction leads
to a significant change in the angle dependence of the tunneling
of two identical Weyl fermions incident from opposite sides of a potential
barrier, it does not affect the HOM interference pattern,
in contrast to previous expectations. We apply our formalism to develop
a Weyl fermion beam-splitter (BS) for controlling the transmission
and reflection coefficients. We calculate the resulting time-resolved correlation
function for two identical Weyl fermions scattering off the BS. 
\end{abstract}
\maketitle
When two indistinguishable bosons are incident on opposite sides
of a 50/50 BS,  Bose-Einstein quantum statistics demands bunching, i.e. the outgoing
bosons must leave together in one of the two outputs, which was first observed with photons in the HOM experiment.\cite{key-1} Observation of zero coincidence for simultaneous
photons is identified by a dip in
the correlation function and rises with
time delay.\cite{key-1} HOM type interference
has been utilized in quantum tests of non-locality\cite{key-2}
and can be used to investigate the degree of indistinguishability of
the incident particles. Also, the HOM experiment is one of the key
elements of linear-optics based quantum computation.\cite{key-3} Several
experiments have already demonstrated the HOM interference
with photons,\cite{key-1,key-4} plasmons,\cite{key-5} levitons,\cite{key-6}
and electrons.\cite{key-7,key-8,key-9} Interestingly, it is possible
to replace the bosons in the HOM interference experiment by fermions,
which leads to the exactly opposite behavior. Due to the Fermi-Dirac
quantum statistics fermions appear in different outputs as identical
fermions have the tendency of antibunching over small distances, leading
to a peak in the coincidence measurement at zero delay. While photons
in vacuum exhibit linear dispersion relation, electrons in gapped
semiconductor materials typically have a quadratic dispersion relation,
which is a major obstacle for observing the fermionic analogue of
the HOM interference due to the spreading of electronic wavefunction.
In order to overcome this obstacle, it is essential to identify physical
systems where the electrons have linear dispersion relation.

One such example is the one-dimensional edge states of quantum Hall
systems exhibiting ballistic conductance and linear dispersion, where the one-dimensional fermionic HOM experiment\cite{key-9}
has been successfully implemented. 
Similar results are expected theoretically for quantum spin Hall states.\cite{key-b}
In order to create a two-dimensional fermionic HOM interference pattern,
we need fermionic particles with a linear dispersion relation in two dimensions. Ideal candidates are Weyl fermions in graphene\cite{key-10,key-11}
and on the surface of 3D topological insulators.\cite{key-12} Here
we show that it is possible to create two-dimensional fermionic HOM interference
pattern by considering the scattering of two Weyl fermions in the
case of a rectangular potential barrier. We show that at specific
incident angles a 50/50 BS for Weyl fermions can be realized, even
when considering the Coulomb interaction between the Weyl fermions.
Interestingly, the Coulomb interaction leads to a substantial change
in angle distribution of the transmission and reflection coefficients.
In Ref. \onlinecite{key-9} a quantum point contact is used as a 50/50 BS
for the electrons. The reduction in the correlation function at zero
time delay is attributed to the Coulomb interaction between the electrons.\cite{key-9}
Here we show that the Coulomb interaction does not affect the correlation
function, i.e. the correlation function is determined solely by the
quantum statistics of the particles.

The realization of fermionic HOM interference experiment is provided
by a three-step process: (i) Generation of single electron
source. (ii) Construction of BS, which is the primary focus
of this work. (iii) Detector for counting the coincidences.
In solid state devices a single electron transistor (SET) can be used
as a source of producing single electrons or a sequential electron
gun.\cite{key-a} The SET consists of a source in the form of a quantum
dot tunnel coupled to a conductor through a quantum point contact.
By applying a sudden voltage step on a capacitively coupled gate,
the charging energy is compensated for and the electron occupying
the highest energy level of the dot is emitted. The final state of
the electron is a coherent wave packet propagating away in the conductor.
Its energy width is given by the inverse tunneling time. 
The absence of an energy gap in 2-D graphene and phenomena related
to Klein tunneling\cite{key-13} make it hard to confine carriers
electrostatically and to control transport on the level of single
particles. However, by focusing on armchair graphene nanoribbons, which are
known to exhibit an energy gap due to boundary conditions,\cite{key-14,key-15,key-16}
this limitation can be overcome. It has been shown that such an energy
gap allows to fabricate tunable graphene nanodevices.\cite{key-17,key-18}
Particularly, in Ref. \onlinecite{key-18} it was shown that quantum dots
in graphene over a size of 100 nm behave as conventional single electron
transistors and exhibit Coulomb blockade. 

\begin{figure}
\includegraphics[width=8cm]{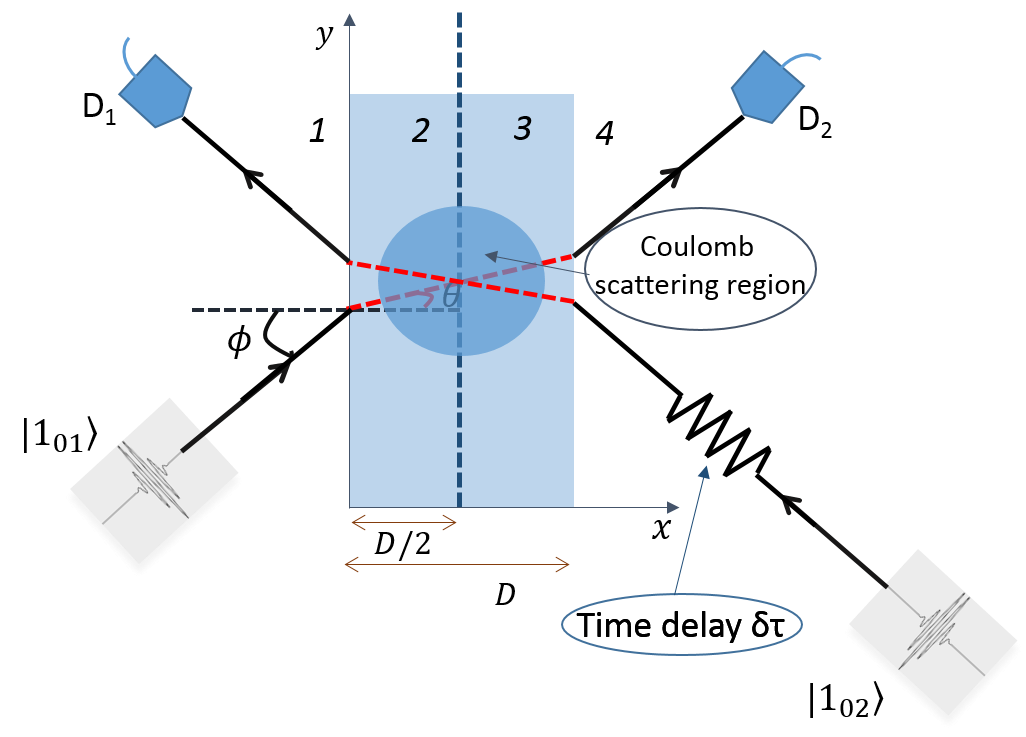}\protect\caption{HOM experiment with Weyl fermion BS.\label{fig:1}}
\end{figure}

It was shown\cite{key-13} that the transmission probability $T$ of Weyl
fermions (in graphene) with energy $E$ through a rectangular potential
barrier of height $V_{0}$ and width $D$ varies as a function of
incident angle $\phi$. 100\% transmission probability is observed
at normal incidence $\phi=0$, a feature known as Klein tunneling. 
Exactly the same result can be obtained for
surface electronic states of 3D topological insulators. The reason
for this coincidence is that in both systems the dynamics of electrons
is defined by similar Hamiltonians. The only difference between the
two systems is that in graphene the pseudo-spin is locked parallel
to the linear momentum and in 3D topological insulators
the real spin is locked perpendicular to linear momentum, respectively,
i.e
\begin{equation}
\mbox{\ensuremath{\hat{H}_{0,g}=v_{F}\mbox{\mbox{\text{\ensuremath{\mathbf{\boldsymbol{\sigma}}}}\ensuremath{\cdot\textbf{\ensuremath{\mathbf{p}}}}}}}}
,\;
\ensuremath{\hat{H}_{0,TI}=\mbox{\ensuremath{v_{eff}\text{\ensuremath{\text{\ensuremath{(\boldsymbol{\sigma}\times\mathbf{p})}}}}}}},\label{eq:1}
\end{equation}
where $\sigma_{i}$'s are Pauli matrices, corresponding to the pseudo-spin
in the case of graphene and to the real spin in the case of 3D topological
insulators, respectively, and $\mathbf{p}$ is the momentum operator.
The angle dependent transmission probability through a potential barrier
can be used to make a BS for Weyl fermions. For observing the HOM
type interference we need to inject two Weyl fermions from the opposite
sides of the barrier as shown in Fig. \ref{fig:1} and their transmissions
and reflections will produce the desired interference. 

We take advantage of the eikonal approximation\cite{key-22} to calculate the phase change
acquired by a Weyl electron when scattering from a second Weyl electron due to the Coulomb interaction. 
We choose the barrier potential height
in such a way that inside the barrier the Coulomb scattering potential
$V(r)$ is small compared to the kinetic energy of the incident
electrons. Although we solve the Coulomb scattering for Weyl fermions
in graphene, our results are general and applicable to surface states
of 3D topological insulators as well. Working in the eikonal approximation
the exact wave function $\Psi$ of the Hamiltonian $H=H_{0}+V(r)$
can be approximated by a semi-classical wave function 
\begin{equation}
\Psi\sim\left(\begin{array}{c}
a\\
b
\end{array}\right)e^{iS(r)/\hbar}.\label{eq:3}
\end{equation}
Starting from the Dirac equation shown in eq. (\ref{eq:1}) and expanding
in powers of $\hbar$, we obtain in zeroth order the relativistic
Hamilton-Jacobi equation 
\begin{equation}
\left|\partial_{x}S(r)\right|^{2}+\left|\partial_{y}S(r)\right|^{2}\approx E^{2}/v_{F}^{2}-2V(r)E/v_{F}^{2}.\label{eq:4}
\end{equation}
We compute $S(r)$ from Eq.~\ref{eq:4} by assuming that the trajectory
is a straight line, which is valid for large energies
and small deflection angles.\cite{key-22}
Eq.~\ref{eq:4} then yields in linear approximation in $V$
\begin{equation}
\frac{S(x)}{\hbar}\approx kx-\frac{1}{\hbar v_{F}}\intop_{-\infty}^{x}2V(b,x')dx'.
\end{equation}
Similar to the non-relativistic derivation,\cite{key-22} we obtain the relativistic scattering amplitude
\begin{equation}
f\left(\mathbf{k},\mathbf{k'}\right)=-i\sqrt{\frac{k}{2\pi}}\intop_{-\infty}^{\infty}dbe^{-ikb\theta}\left[e^{2i\triangle(b)}-1\right],\label{eq:7}
\end{equation}
where
$
\triangle(b)=-\frac{1}{2\hbar v_{F}}\intop_{-\infty}^{\infty}dx'V(b,x')
$
and $\theta$ is the angle between $\mathbf{k}\mbox{ and }\mathbf{k'}$.
Eq. \ref{eq:7} is in agreement with the optical theorem in scattering
theory.\cite{key-22} Eq. (\ref{eq:7}) can be solved
for the screened Coulomb potential, i.e. the Yukawa potential with
$V(b,x')=U_{0}\exp\left(-\mu\sqrt{b^{2}+x^{2}}\right)/\mu\sqrt{b^{2}+x^{2}}$,
where $\mu^{-1}$ is the screening length, for graphene $\mu=g_{s}g_{v}e^{2}k_{F}/\kappa\hbar v_{F}$,
$\kappa$ is the background lattice dielectric constant, $U_{0}=e^{2}\mu/4\pi\kappa\epsilon_{0}$,
and $k_{F}$ is the Fermi wave vector. 
In the lab frame $\theta\longrightarrow\text{\ensuremath{\theta/2}}$. The
phase change $\Delta$ in the forward direction acquired by the particle
while passing through the scattering region can be evaluated by setting $\left|\mathbf{k}\right|=\left|\mathbf{k'}\right|=k_{F}$
for elastic scattering, i.e.
\begin{equation}
\text{\ensuremath{\Delta=\underset{\theta\longrightarrow0}{\lim}Re\left(\sqrt{k_{F}}f\left(\mathbf{k},\mathbf{k'}\right)\right)=\mbox{\ensuremath{-\frac{\sqrt{2\pi}U_{0}}{\hbar v_{F}\mu}\frac{k_{F}}{\mu}}}}}.\label{eq:10}
\end{equation}
It is now straightforward to solve the tunneling problem shown in
Fig. \ref{fig:1}. The electron is incident on the barrier from right
at an angle $\phi$ with respect to the $x$ axis. It propagates at an angle $\theta$ in region 2
and is transmitted in region 3 at the same angle $\phi$. 
Using the notation in Ref. \onlinecite{key-13}, the components of the Weyl
spinor $\Psi_{1}$ and $\Psi_{2}$ can be written as $\Psi_i(x,y)=\Psi_i(x)e^{ik_{y}y}$, $i=1,2$, with
\begin{eqnarray}
\Psi_{1}(x) & = & \begin{cases}
e^{ik_{x}x}+re^{-ik_{x}x} & x<0\\
ae^{iq_{x}x}+be^{-iq_{x}x} & 0<x<\frac{D}{2}\\
ae^{iq_{x}x+i\Delta}+be^{-iq_{x}x-i\Delta} & \frac{D}{2}<x<D\\
te^{ik_{x}x+i\Delta} & x>D
\end{cases},
\\
\Psi_{2}(x) & = & \begin{cases}
s\left[e^{ik_{x}x+i\phi}-re^{-ik_{x}x-i\phi}\right] & x<0\\
s'\left[ae^{iq_{x}x+i\theta}-be^{-iq_{x}x-i\theta}\right] & 0<x<\frac{D}{2}\\
s'\left[ae^{iq_{x}x+i\theta+i\Delta}-be^{-iq_{x}x-i\theta-i\phi}\right] & \frac{D}{2}<x<D\\
ste^{ik_{x}x+i\phi+i\Delta} & x>D
\end{cases},
\end{eqnarray}
where $k_{x}=k_{F}\cos\text{\ensuremath{\phi,}\ensuremath{k_{y}=k_{F}\sin\text{\ensuremath{\phi}}}}$ are
the components of the wavevector outside the barrier and $\ensuremath{q_{x}=\sqrt{\mbox{\ensuremath{\left(E-V_{0}\right)^{2}/\left(\hbar v_{F}\right)^{2}}}-k_{y}^{2}}}$
and $\tan\theta=k_{y}/q_{x}$. The transmission coefficient$t$ can
be evaluated by using the continuity conditions at $x=0$ and $x=D$
and is
\begin{eqnarray}
t & = & 2\exp(-ik_{x}D)\cos\theta\cos\phi /\left\{ss'\left[e^{-i\left(q_{x}D+\Delta\right)}\cos\left(\theta+\phi\right) \right.\right. 
\nonumber\\
& & +\left.\left.e^{i\left(q_{x}D+\Delta\right)}\cos\left(\theta-\phi\right)\right]-2i\sin\left(q_{x}D+\Delta\right)\right\}.\label{eq:a1}
\end{eqnarray}

\begin{figure}
\includegraphics[width=6cm]{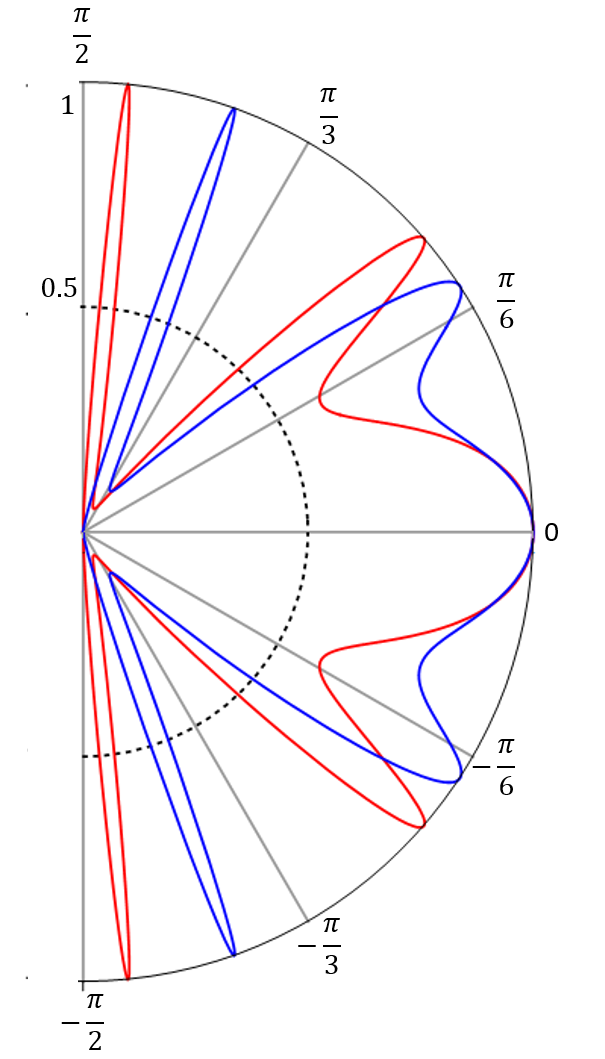}\protect\caption{Transmission probability $T$ as a function of incident angle $\phi$.
The electron concentration $n$ outside the barrier is chosen as $0.5\times10^{12}$
$\mbox{cm}{}^{-2}$. This corresponds to a Fermi energy and wavelength of incident electrons of $E_{F}\approx 80$ meV
and $\lambda\approx 50$ nm, respectively. The barrier height $V_{0}=200$ meV. The red curve is the
solution for $\text{\textgreek{D}}=0$ and the blue curve is the solution
for $\text{\textgreek{D}}=-0.63$. Black (dashed) semicircle is drawn
at 50\% transmission probability.\label{Fig3}}
\end{figure}

In Fig. \ref{Fig3} the transmission coefficient $T=t^*t$ is plotted
as a function of incident angle $\phi$ for the cases when $\text{\textgreek{D}=0 }$(red
curve) and $\Delta=-\sqrt{2\pi}U_{0}k_{F}/\hbar v_{F}\mu^{2}$ (blue
curve). Interestingly, the Coulomb interaction results in a substantial
shift of the transmission peaks while preserving Klein tunneling.
In the limit $V_{0}\ll E$ ,
\begin{equation}
T=\frac{cos^{2}\text{\ensuremath{\phi}}}{1-cos^{2}(q_{x}D\text{+\ensuremath{\Delta}})sin^{2}\text{\ensuremath{\phi}}}.\label{eq:11}
\end{equation}
For normal incidence $T$ is always $1$, regardless of the height
and width of the barrier. Away from normal incidence, the other transmission
peaks correspond to the condition of constructive interference, which
occurs when $q_{x}D\text{+\ensuremath{\text{\textgreek{D}}}}=n\pi$
where $n=0,\pm1,\pm2,....$. Comparing Eq. (\ref{eq:11}) with the result
in Ref. \onlinecite{key-13}, there is an additional phase $\text{\ensuremath{\Delta}}$
in the denominator, which comes from the Coulomb interaction. It can
be seen from Fig. \ref{Fig3} that at certain angles the transmission
coefficient is 50\%. For these angles of incidence this modified barrier
can be used as a 50/50 BS. At the same $\phi$, the Coulomb interaction then leads to an asymmetry in $T$ and $R$. In addition, we can
change the transmission and reflection coefficients to any desired
value ranging between 0 and 1 by tuning $\phi$. 

\begin{figure}
\includegraphics[width=8cm]{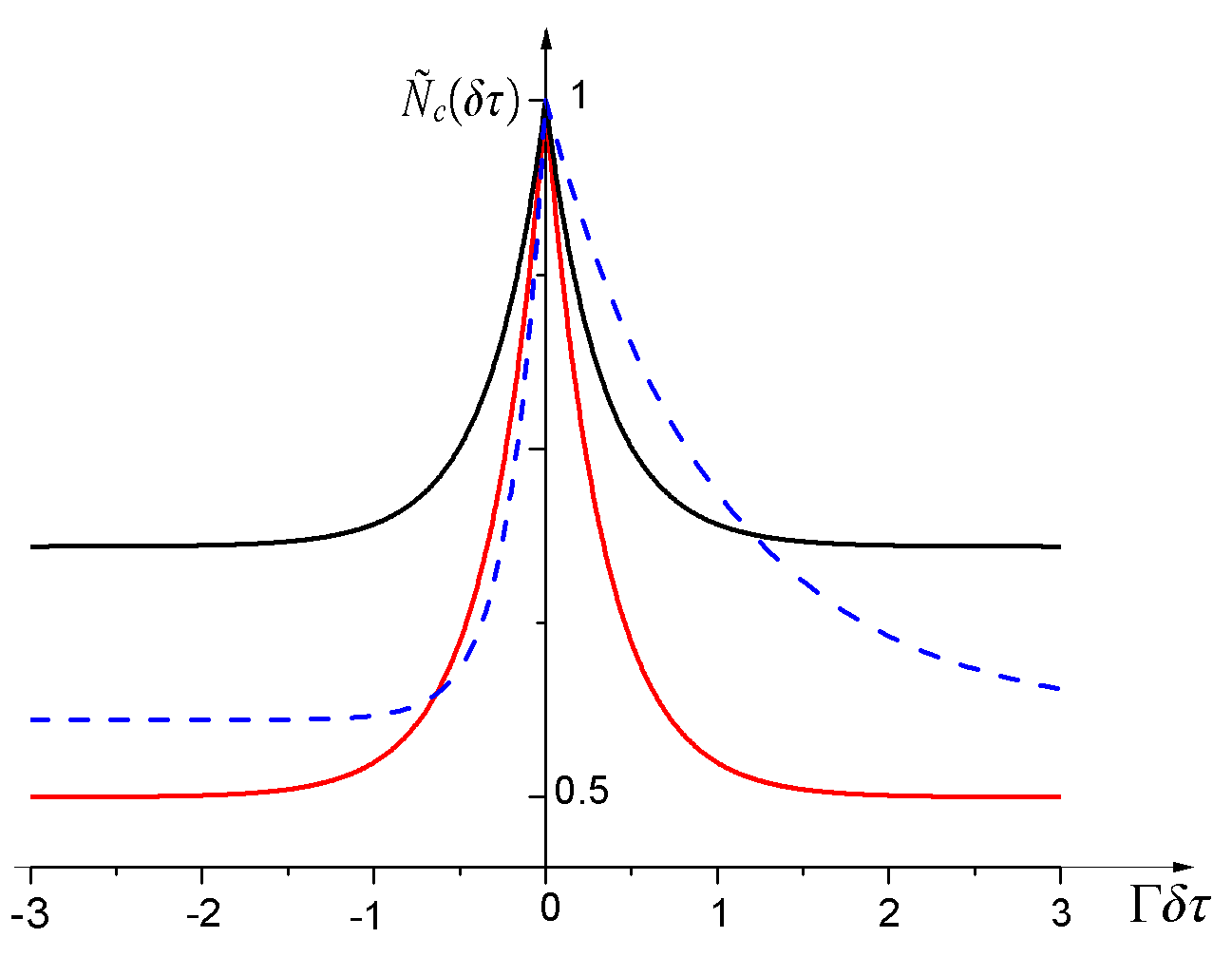}\protect\caption{Interference peak for normalized number of coincidences $\widetilde{N}_{c}(\delta\text{\textgreek{t}})$
against time delay $\delta\text{\textgreek{t}}$. Red curve is for
R=T=1/2 and for $\Gamma_{1}=\Gamma_{2}=\Gamma$, black curve is for
$R=1/5,T=4/5$ and for $\Gamma_{1}=\Gamma_{2}=\Gamma$. Blue curve
is for $T=1/3,R=2/3$ and$\Gamma_{1}=5\Gamma/3,\Gamma_{2}=\Gamma/3$,
where $\Gamma=10^{-12}s^{-1}$.\label{fig:5}}
\end{figure}

The schematic diagram of the HOM experiment is shown in Fig. \ref{fig:1}.
It consists of two SET's as the sources of the two electrons, a BS
(orange line) and electron counters (blue pentagons).\cite{key-23,key-24}
The BS is considered to be lossless, i.e. $T+R=1$.
Let us now consider two Weyl fermions that are incident on the BS
from opposite sides. Let $\tau_{1}$ be the time it takes for the electrons
to get from the source to the detector. We define $\delta\tau$
as the time delay between the two incident electrons. $\delta\tau$
can be introduced either by displacing the position of the BS towards
one of the sources or by introducing the time delay between the switching
pulses of the two SET's. Our goal is to calculate the correlation
function corresponding to the coincidence counts at the two detectors
as a function of the time delay $\delta\tau$.
The inputs of the BS are described by the indices $01$,$02$ i.e.
$c_{01}^{\dagger}\left|0_{01},0_{02}\right\rangle =\left|1_{01},0_{02}\right\rangle $and
$c_{02}^{\dagger}\left|0_{01},0_{02}\right\rangle =\left|0_{01},1_{02}\right\rangle $,
where $c_{01}^{\dagger}(c_{01})$ are electron creation (annihilation)
operators. We omit the spin index because we assume that the two electrons
have parallel spins. Similarly, the outputs are described by the indices
$1,2.$ The output operators are related to the input operators through
the following linear scattering relations
\begin{eqnarray}
\hat{c}_{1}(t) & = & \sqrt{T}\hat{c}_{01}(t-\tau_{1})+i\sqrt{R}\hat{c}_{02}(t-\tau_{1}+\delta\tau),\label{eq:12}
\\
\hat{c}_{2}(t) & = & \sqrt{T}\hat{c}_{02}(t-\tau_{1})+i\sqrt{R}\hat{c}_{01}(t-\tau_{1}-\delta\tau),\label{eq:13}
\end{eqnarray}
where $i$ corresponds to a $\pi/2$ phase shift and $\hat{c}_{0j}(t)=\xi_{j}(t)\hat{c}_{02}$.
$\xi_{j}(t)$ is the distribution function in time. Electrons emitted
from the SET usually follow an exponential profile in time, i.e. $\xi_{j}(t)=\Theta(t)\exp(-\Gamma_{j}t/2)\exp(i\omega t)$.\cite{key-9}
$\Theta(t)$ is the Heavyside step function and $\Gamma_{j}$is
the SET emission rate of the electron. The correlation function describing
the joint probability of detection of electrons at the two detectors
at times $t$ and $t+\tau$ is
\begin{equation}
P_{12}(t)=C\left\langle 0\left|\hat{c}_{02}\hat{c}_{01}\hat{c}_{1}^{\dagger}(t)\hat{c}_{2}^{\dagger}(t+\tau)\hat{c}_{2}(t+\tau)\hat{c}_{1}(t)\hat{c}_{01}^{\dagger}\hat{c}_{02}^{\dagger}\right|\text{0}\right\rangle .\label{eq:14}
\end{equation}
$C$ is the normalization constant. This can readily be evaluated
by means of Eqs. (\ref{eq:12}) and (\ref{eq:13}).
The number of coincidence counts $N_{c}(1,2)$ can be obtained by
integrating $P_{12}(t)$ over time $t$. This yields
\begin{eqnarray}
& & \frac{N_{c}(\delta\tau)}{C}\Gamma_{1}\Gamma_{2} = \widetilde{N}_{c}(\delta\tau)
=T^{2}+R^{2}+RT\frac{8\Gamma_{1}^{2}\Gamma_{2}^{2}}{\left(\Gamma_{1}+\Gamma_{2}\right)^{2}} 
\nonumber\\
\lefteqn{\times\left\{ \exp(\Gamma_{1}\delta\tau)\Theta(-\delta\tau)
+\exp(-\Gamma_{2}\delta\tau)\Theta(\delta\tau)\right\} ,}\label{eq:17}
\end{eqnarray}
where $\widetilde{N}_{c}(\delta\tau)$ is the normalized
number of coincidences. Eq. (\ref{eq:17}) is our main result. The
coincidence counts depend both on the time delay $\delta\tau$
and the transmission and reflection coefficients. The coincidence
counts can be tuned by introducing an asymmetry in the reflection
and transmission coefficients. For perfect transmissions and reflections
$\widetilde{N}_{c}(\delta\tau)$ remains at unity
regardless of the value of $\delta\tau$.
For large $\delta\tau$ the third term on the right
hand side of Eq. (\ref{eq:17}) goes to zero, and the expression for
the coincidence counts reduces to $T^{2}+R^{2}$. In case of identical
electron sources, i.e. $\Gamma_{1}=\Gamma_{2}$, Eq. (\ref{eq:17}) can
be simplified to 
\begin{eqnarray}
\widetilde{N}_{c}(\delta\tau) & = & T^{2}+R^{2}+2RT\left\{ \exp(\Gamma_{1}\delta\tau)\Theta(-\delta\tau) \right.
\nonumber\\
& & +\left.\exp(-\Gamma_{2}\delta\tau)\Theta(\delta\tau)\right\}. \label{eq:18}
\end{eqnarray}
Note that for $\delta\tau=0$ $\widetilde{N}_{c}(\delta\tau)=\left(T+R\right)^{2}=1$,
no matter what the values of $T$ and $R$ are, which reflects the
antibunching of fermions. In Fig. \ref{fig:5} we plot the coincidence
counts for different $R$ and $T$ and for different
values of $\Gamma$'s (blue) as a function of the time delay $\delta\tau$.
Note that, in contrast to the expectation in Ref. \onlinecite{key-9},
the Coulomb interaction does not reduce the peak at $\delta\tau=0$. 

In conclusion, we developed the theoretical model of the two-dimensional
HOM type interference with Weyl fermions in graphene and in 3D topological
insulators. The two-dimensional setup allows for the tuning of the
transmission and reflection coefficients by varying the angle of incidence
of the two Weyl fermions. We provide the description a realistic BS
for Weyl fermions, including the effects of Coulomb interaction. Our
results show that the Coulomb interaction does not affect the fermionic HOM
peak (Pauli peak) for Weyl fermions within the eikonal approximation.
We conjecture that as long as the detectors can absorb electrons laterally spread by the Coulomb interaction,
our results are valid beyond the eikonal approximation.

Acknowledgments. We acknowledge support from NSF (grant ECCS-0901784), AFOSR (grant FA9550-09-1-0450), and NSF (grant ECCS-1128597).

\end{document}